\documentclass{article}
\usepackage[utf8]{inputenc}
\usepackage{comment}
\usepackage{cite}
\usepackage{graphicx,amsmath,amsfonts,amssymb
}
\usepackage{color}

\begin{document}
\begin{titlepage}
\begin{center}


\vskip .5in

{\Large \bf
Stochastic gravitational waves associated with \\the formation of primordial black holes 

}
\vskip .45in

{\large
Tomohiro Nakama$^{1}$,
Joseph Silk$^{1,2,3}$
and Marc Kamionkowski$^{1}$
}

\vskip .45in%
{\em
$^1$
   Department of Physics and Astronomy, \\ Johns Hopkins University, Baltimore, Maryland 21218,
USA}\\
\vskip .05in
{\em
$^2$
   Institut d'Astrophysique de Paris, UMR 7095, CNRS, \\UPMC Universit\'e Paris VI, 98 bis Boulevard Arago, 75014 Paris,
	France
    }\\
    \vskip .05in
    {\em
$^3$
   BIPAC, Department of Physics, \\University of Oxford, Keble Road, Oxford OX1 3RH, United Kingdom
    }

\end{center}

\vskip .4in
\begin{abstract}
Primordial black hole (PBH) mergers have been proposed as an explanation  for the gravitational wave events detected by the LIGO collaboration.  Such PBHs may be formed in the early Universe as a result of the collapse of extremely rare high-sigma peaks of primordial fluctuations on small scales, as long as the amplitude of primordial perturbations on small scales is enhanced  significantly relative to the amplitude of perturbations observed on large scales.  One consequence of these small-scale perturbations is generation of stochastic gravitational waves  that arise at second order in scalar perturbations,  mostly before the formation of the PBHs.  These induced gravitational waves have been shown, assuming gaussian initial conditions, to be comparable to the current limits from the European Pulsar Timing Array, severely restricting this scenario. We show, however, that models with enhanced fluctuation amplitudes typically involve non-gaussian initial conditions.  With such initial conditions, the current limits from pulsar timing can be evaded. The amplitude of the induced gravitational-wave background can be larger or smaller than the stochastic gravitational-wave background  from supermassive black hole binaries.
\end{abstract}

\end{titlepage}

\section{Introduction}
The LIGO collaboration recently detected gravitational waves (GWs) from merging black holes (BHs) \cite{Abbott:2016blz,TheLIGOScientific:2016wfe,Abbott:2016nmj,Abbott:2016nhf,TheLIGOScientific:2016pea}, and the first event of these originated from BHs with masses of $\sim\, 30M_\odot$.
Although there are a number of stellar-astrophysical origins for these BHs \cite{Hosokawa:2015ena,Rodriguez:2016kxx,Woosley:2016nnw,Chatterjee:2016hxc,deMink:2016vkw,Hartwig:2016nde,Inayoshi:2016hco,Rodriguez:2016avt,Kovetz:2016kpi}, it is also possible  \cite{Bird:2016dcv,Sasaki:2016jop,Clesse:2016vqa} that they may be primordial black holes (PBHs) \cite{Hawking:1971ei,1966AZh....43..758Z,Carr:1974nx,Carr:1975qj}.  A relatively large population of PBHs might have been formed as a result of a spike in the spectrum of the initial curvature perturbations  \cite{Ivanov:1994pa,Yokoyama:1995ex,Bullock:1996at,Yokoyama:1998pt,Kawasaki:1998vx,Kawasaki:2006zv,Saito:2008em,Bugaev:2008bi,Cheng:2016qzb,Kawasaki:2016pql},  which leads to the presence of extremely rare peaks collapsing to PBHs during, e.g., the radiation-dominated era.  Various tests of this early-Universe scenario for PBHs have been discussed in \cite{Clesse:2016ajp,Chen:2016pud,Nakamura:2016hna,Cholis:2016kqi,Brandt:2016aco,Raccanelli:2016cud,Munoz:2016tmg,Namikawa:2016edr,Carr:2016drx,Gaggero:2016dpq,Schutz:2016khr,Kashlinsky:2016sdv}. \textcolor{black}{References such as} ~\cite{Carr:2009jm,Carr:2016hva} provide additional observational constraints on PBHs in general.

One of the interesting implications of this scenario for PBHs is the generation of a stochastic GW background at around their formation epoch (in addition to any stochastic background \textcolor{black}{of higher frequencies} from the mergers of PBH binaries during the epoch of matter domination \cite{Mandic:2016lcn,Wang:2016ana}).  The formation of a sufficient number of PBHs would require, however \cite{Carr:1994ar,Kim:1996hr,Green:1997sz,Zaballa:2006kh,Bugaev:2008gw,Josan:2009qn,Josan:2010cj,Kohri:2014lza}, that the amplitude of primordial perturbations be considerably enhanced on small scales relative to that observed on large scales \cite{Ade:2015xua}.
In this case, a stochastic GW background of relatively large amplitude is generated at second order in scalar perturbations when they are on superhorizon scales (induced GWs) \cite{Matarrese:1997ay,Carbone:2004iv,Ananda:2006af,Baumann:2007zm,Assadullahi:2009jc,Saito:2008jc,Saito:2009jt,Bugaev:2009zh,Bugaev:2010bb,Alabidi:2012ex,Alabidi:2013wtp,Choudhury:2013woa,Kawasaki:2013xsa,Saga:2014jca}.
Furthermore, primordial fluctuations could lead to the formation of shocks, from which GWs are also emitted \cite{Pen:2015qta}. This also leads to a large-amplitude GW background if the RMS fluctuations are sufficiently large.
For either induced GWs or GWs from shocks, a \textit{global} generation of GWs is considered; that is, GWs are emitted everywhere in the Universe, even though overdensities collapsing to PBHs are extremely rare (see, e.g, \cite{Carr:2009jm}). In either case, the amplitudes of these global GWs are determined by the amplitudes of primordial fluctuations.  The frequency of GWs associated with the formation of PBHs of $\sim 30\,M_\odot$ is \textcolor{black}{observed to be} roughly $\sim \mathrm{nHz}$ \textcolor{black}{due to the cosmic expansion}, and the predicted amplitude is already comparable to the current limits set by pulsar timing arrays, thus excluding (apparently) the PBH scenario for the initial LIGO event, or at least the small-scale-spike scenario for production of such PBHs.

In prior work, however, primordial fluctuations have been assumed to be gaussian.  As we discuss below, however, the inflationary physics that generates a spike in the power spectrum for primordial perturbations is also likely to induce non-gaussianity on these small scales \cite{Bullock:1996at,Bullock:1998mi}.
Hence it is important to investigate the effects of non-gaussianity on the generation of global GWs.  To this end, we use some simple models of non-gaussianity to illustrate how the GW amplitude can change.  If we discard the assumption of gaussianity, then the power-spectrum amplitude required to produce the same abundance of PBHs can be changed and, in particular, reduced.  As an extreme example, it has been shown \cite{Nakama:2016kfq} that the amplitude can be kept as small as $\sim 10^{-4}$ of the standard scale-invariant spectrum in multi-field inflation models in \cite{Nakama:2016kfq}.  There the possibility to explain high-redshift supermassive black holes by PBHs formed by the collapse of primordial perturbations was explored, without causing unacceptably large $\mu$ distortions to the frequency spectrum of the cosmic microwave background.
The suppression we explore here of the power-spectrum amplitude and resulting GW amplitude by non-gaussianity is quite similar.  We focus explicitly on induced GWs, but note \cite{Pen:2015qta} that the amplitude of GWs from shocks should be similar, but at somewhat lower frequency. 

This paper is constructed as follows. In \S II we revisit the relation between the mass of PBHs and the typical frequency of GWs associated with their formation. In \S III we discuss the dependence of induced GWs on non-gaussianity using several models for non-gaussianity, comparing induced GWs with current pulsar timing limits, a prediction for GWs from supermassive black hole binaries and also for the future sensitivity of the Square Kilometer Array, and we conclude in \S IV.

\section{Mass-frequency relation revisited}
As will be shown, the relation between the mass of PBHs and the typical frequency of GWs associated with their formation is crucial to determine whether they are probed by pulsar timing.  Hence we revisit this issue closely in the following.

The mass of PBHs can be approximately evaluated by the horizon mass at the moment when the fluctuations reenter the horizon. In cosmology, the horizon reentry  is usually defined by $k=aH$, and  let us therefore write
\begin{equation}
M=\gamma \frac{4\pi}{3}(\rho_r H^{-3})_{k=aH},\label{mass}
\end{equation}
where $\gamma={\cal O}(1)$ parameterizes the deviation from the horizon mass. On the other hand, in numerical simulations the horizon  reentry has often been defined by the moment when the radius 
of the overdensity coincides with the Hubble radius \cite{Musco:2004ak,Polnarev:2006aa,Musco:2008hv,Musco:2012au,Polnarev:2012bi,Nakama:2013ica}, and hence let us introduce $\gamma'$ to write 
\begin{equation}
M=\gamma'\frac{4\pi}{3}(\rho_r H^{-3})_{ar=H^{-1}}.
\end{equation}
How best to relate these two definitions of horizon reentry is not trivial. If one simply uses $r=k^{-1}$, then both definitions coincide ($\gamma=\gamma'$), but it might be more reasonable to assume the radius is one-fourth of the wavelength: $r=\lambda/4=(\pi/2)k^{-1}$. Suppose one assumes $r=\alpha k^{-1}$, 
then since $t(ar=H^{-1})=\alpha^2 t(k=aH)$ and $\rho_r H^{-3}\propto H^{-1}\propto t$, $\gamma=\alpha^2 \gamma'$.

There have been attempts to calculate the masses of PBHs by numerical simulations. 
One of the difficulties in  determining the masses of PBHs by numerical simulations is the appearance of a singularity, and a simulation has to be terminated before the mass asymptotes to a constant value if one adopts a slicing facing the singularity, such as the one used in \cite{1979STIN...8010983N}. 
One way to circumvent this problem is to use what is called null (or observer time) slicing \cite{Hernandez:1966zia,1995ApJ...443..717B,0264-9381-6-2-012,Musco:2004ak,Polnarev:2006aa,Musco:2008hv,Musco:2012au,Nakama:2013ica,Nakama:2014fra} under spherical symmetry, in which the spacetime is sliced along null geodesics of hypothetical photons emitted from the center and escaping the formed BH horizon to reach spatial infinity, as illustrated in Fig.~8 of Ref.~\cite{Nakama:2013ica}. With this slicing, one can follow the formation of the BH horizon and subsequent accretion until the mass converges, as is shown in, e.g., Fig.~9 of Ref.~\cite{Nakama:2013ica}. 

If the amplitude is extremely close to threshold, 
the perturbation enters into a 
quasi-stable, relatively long-lived intermediate state, during which a substantial fraction of matter is blown away from the central overdensity as a relativistic wind \cite{Musco:2008hv}. If eventually the perturbation collapses to form a PBH from an intermediate state, its mass can be substantially smaller than the horizon mass ($\gamma'\ll 1$), and the resultant mass follows a scaling relation of the form $M\propto (\delta-\delta_{\mathrm{th}})^\gamma$ (critical phenomenon) \cite{Niemeyer:1999ak,Hawke:2002rf,Musco:2008hv,Musco:2012au}. 
In Ref.~\cite{Niemeyer:1999ak}, the critical phenomenon was investigated using  null slicing, 
but they provide initial conditions at around horizon reentry and hence they are  contaminated by unrealistic decaying modes as noted in  \cite{Shibata:1999zs}. 
Indeed, in Ref.~\cite{Shibata:1999zs}, another slicing which also avoids the singularity was used focusing on growing modes, and they found that the mass depends on the profile and amplitude, but their simulations were terminated before the mass fully converged. 

We assume that most PBHs form from peaks sufficiently above the threshold, so that $\gamma'$ is not significantly smaller than unity.\footnote{This is the case at least if primordial fluctuations follow gaussian statistics 
\cite{Niemeyer:1997mt,Yokoyama:1998xd}. Though 
the probability density is larger at threshold than at larger amplitudes since the probability density is rapidly decreasing there, the amplitude has to be extremely close to the threshold in order for the critical phenomenon to be important, and also PBH masses are small for such cases and hence they are relatively unimportant.
The critical phenomenon 
would be relatively more important when the probability density decreases more slowly around the threshold than in the gaussian case.  
}
According to Ref.~\cite{Musco:2008hv} $\gamma'\gtrsim 1$ for $(\delta-\delta_{\mathrm{th}})\gtrsim 0.02$ for their initial conditions. 
An example shown in Fig.~9 of Ref.~\cite{Nakama:2013ica} 
corresponds to a compensated\footnote{Uncompensated profiles
would probably lead to larger masses
\cite{1979ApJ...232..670B,Shibata:1999zs}.}
\cite{Harada:2015yda}, gaussian-like initial profile, and also
to a growing mode. In addition this case is sufficiently above
the threshold and not affected by the critical phenomenon. 
For this example $\gamma'\simeq 1.2$, which translates to $\gamma\simeq 3.$, if we use $\alpha=\pi/2$. 

Next, we rederive the relation between the mass of PBHs and the wavenumber of fluctuations collapsing to them, and also the typical frequency of GWs associated with their formation, leaving $\gamma$ of (\ref{mass}) unspecified.
In this context, the horizon mass during the radiation-dominated era is related to that at the moment of the matter-radiation equality:
\begin{equation}
M_{\mathrm{eq}}=\frac{4\pi}{3}2\rho_{\mathrm{eq}}H_{\mathrm{eq}}^{-3}=\frac{8\pi\Omega_r\rho_{\mathrm{cr}}}{3a_{\mathrm{eq}}k_{\mathrm{eq}}^3}\simeq 2.8\times 10^{17}M_\odot,
\end{equation}
where $\rho_{\mathrm{cr}}=3H_0^2/8\pi G$ and we have used $\Omega_{r}\simeq 8\times 10^{-5}$ \cite{Weinberg:2008zzc}, $z_{\mathrm{eq}}\simeq 3400$, $k_{\mathrm{eq}}\simeq 0.01\mathrm{Mpc}^{-1}$ and $H_0\simeq 67 \mathrm{km}\mathrm{s}^{-1}\mathrm{Mpc}^{-1}$\cite{Ade:2015xua}.
Then the mass of PBHs is related to the temperature as
\begin{equation}
\frac{M}{M_{\mathrm{eq}}}=\gamma\left(\frac{\rho_r}{2\rho_{\mathrm{eq}}}\right)^{-1/2}=\sqrt{2}\gamma\left(\frac{g}{g_{\mathrm{eq}}}\right)^{-1/2}\left(\frac{T}{T_{\mathrm{eq}}}\right)^{-2}, \quad M\simeq1.5\times 10^5\gamma M_\odot\left(\frac{g}{10.75}\right)^{-1/2}\left(\frac{T}{1\mathrm{MeV}}\right)^{-2},
\end{equation}
where $g$ is the number of degrees of freedom of relativistic species, and
$T_{\mathrm{eq}}\simeq z_{\mathrm{eq}}T_0\simeq 9300\mathrm{K} \:(0.80\mathrm{eV)}$ with $T_0\simeq 2.726 \mathrm{K} $\cite{Fixsen:2009ug}.
For $1\mathrm{MeV}\lesssim T\lesssim 100\,\mathrm{MeV}$, $g\simeq 10.75$ (see, e.g., Ref.~\cite{Husdal:2016haj} for a recent calculation), and this range
 of temperatures corresponds to PBH masses of
$15\,M_\odot\lesssim M\lesssim 1.5\times 10^5\,M_\odot$.
Since the temperature of radiation evolves according to the conservation of entropy, i.e., $ga^3T^3=\mathrm{const}.$ and hence $\rho_r\propto gT^4\propto g^{-1/3}a^{-4}$.\footnote{Strictly speaking, \textcolor{black}{the numbers of} degrees of freedom defined in terms of entropy and energy  should be distinguished, and they start to slightly deviate at around the epoch of electron-positron annihilation: the former becomes $\simeq 3.9$ whereas the latter becomes $\simeq 3.4$ \cite{Husdal:2016haj}. Since the difference is not significant and also the dependence on $g$ is relatively weak, we use $g_{\mathrm{eq}}=3.5$.
} Using this relation we find
\begin{equation}
\frac{M}{M_{\mathrm{eq}}}=\frac{\gamma}{2}\frac{\rho}{\rho_{\mathrm{eq}}}\left(\frac{H_{\mathrm{eq}}}{H}\right)^3=\frac{\gamma}{2}\left(\frac{g_{\mathrm{eq}}}{g}\right)^{1/3}\left(\frac{k_{\mathrm{eq}}}{k}\right)^2\frac{a_{\mathrm{eq}}^2H_{\mathrm{eq}}}{a^2H}.
\end{equation} 
Since $a_{\mathrm{eq}}^2H_{\mathrm{eq}}/a^2H=\sqrt{2}(g/g_{\mathrm{eq}})^{1/6}$,
\begin{equation}
M=\frac{\gamma }{\sqrt{2}}M_{\mathrm{eq}}\left(\frac{g_{\mathrm{eq}}}{g}\right)^{1/6}\left(\frac{k_{\mathrm{eq}}}{k}\right)^2\simeq 17\gamma M_\odot \left(\frac{g}{10.75}\right)^{-1/6}\left(\frac{k}{10^6\mathrm{Mpc}^{-1}}\right)^{-2}.\label{massk}
\end{equation}
This can be inverted as
\begin{equation}
k=\left(\frac{\gamma}{\sqrt{2}}\right)^{1/2}k_{\mathrm{eq}}\left(\frac{g_{\mathrm{eq}}}{g}\right)^{1/12}\left(\frac{M_{\mathrm{eq}}}{M}\right)^{1/2}\simeq 7.5\times 10^5\gamma^{1/2}\mathrm{Mpc}^{-1}\left(\frac{g}{10.75}\right)^{-1/12}\left(\frac{M}{30M_\odot}\right)^{-1/2},
\end{equation}
or in terms of frequency $f=ck/2\pi$,
\begin{equation}
f\simeq 1.2\gamma^{1/2}\mathrm{nHz}\left(\frac{g}{10.75}\right)^{-1/12}\left(\frac{M}{30M_\odot}\right)^{-1/2}.\label{freq}
\end{equation}
That is, the typical frequency of GWs associated with the formation of PBHs of $\sim 30M_\odot$ is \textcolor{black}{observed to be} $\sim$ nHz, \textcolor{black}{due to the cosmic expansion,} and hence they are likely to be probed by pulsar timing, as  will be shown below.

Instead of $M_{\mathrm{eq}}$, one can also relate the mass of PBHs to the current horizon mass $M_0=4\pi\rho_{\mathrm{cr}}/(3H_0)^3\simeq 4.7\times 10^{22}\,M_\odot$ as follows. 
From the conservation of entropy, 
\begin{equation}
\rho_r=\left(\frac{g_0}{g}\right)^{1/3}a^{-4}\Omega_r\rho_{\mathrm{cr}},
\end{equation}
then, 
\begin{equation}
M=\frac{4\pi\gamma}{3a}k^{-3}\left(\frac{g_0}{g}\right)^{1/3}\Omega_r\rho_{\mathrm{cr}}.
\end{equation}
The scale factor here can be eliminated by
\begin{equation}
H^2=\left(\frac{k}{a}\right)^2=\frac{8\pi G}{3}\left(\frac{g_0}{g}\right)^{1/3}a^{-4}\Omega_r\rho_{\mathrm{cr}}=\left(\frac{g_0}{g}\right)^{1/3}a^{-4}\Omega_rH_0^2,
\end{equation}
\begin{equation}
a^{-1}=k\left(\frac{g}{g_0}\right)^{1/6}\Omega_r^{-1/2}H_0^{-1}.
\end{equation}
Then
\begin{equation}
M=\frac{4\pi\gamma}{3}\left(\frac{g_0}{g}\right)^{1/6}\frac{\Omega_r^{1/2}\rho_{\mathrm{cr}}}{H_0k^2}
=\gamma \Omega_r^{1/2}M_0\left(\frac{g_0}{g}\right)^{1/6}\left(\frac{H_0}{k}\right)^2.
\end{equation}

\section{Dependence of induced GWs on non-gaussianity}
\begin{figure}[htp]
\begin{center}
\includegraphics[width=10cm,keepaspectratio,clip]{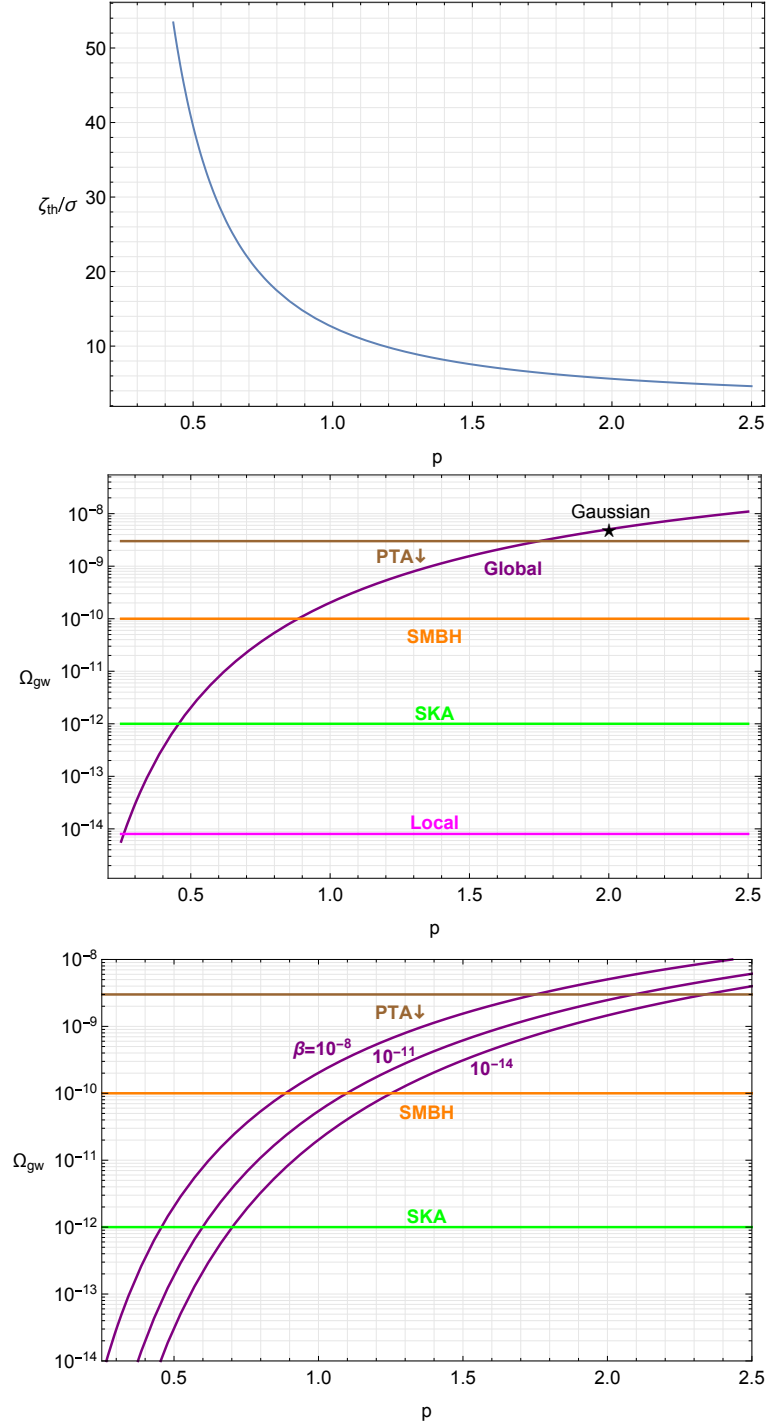}
\end{center}
\caption{
Top: 
The ratio $\zeta_{\mathrm{th}}/\sigma$ as a function of $p$, controlling deviation from a gaussian PDF in Eq.~(\ref{pdf2}), with $\beta=10^{-8}$. When $p=2$ the PDF is gaussian, for which $\zeta_{\mathrm{th}}/\sigma\simeq 6$. 
Center: 
Dependence of global or induced GWs on 
$p$. 
Local gravitational waves due to non-spherical collapse are also shown, which do not depend on non-gaussianity (but see the footnote 7). Current limits from European Pulsar Timing Array \cite{Lentati:2015qwp} (PTA), 
a prediction of stochastic gravitational waves from supermassive black hole (SMBH) binaries
and a predicted sensitivity of the Square Kilometer Array (SKA) \cite{Kramer:2010tm} are also shown.
Bottom: 
The dependence of induced GWs on $\beta$. 
}
\label{main}
\end{figure}

Let us use the following class of non-gaussian probability density functions (PDFs), introduced in Ref.~\cite{Nakama:2016kfq}\footnote{
For general $p$, derivatives at $\zeta=0$ are discontinuous and hence 
this PDF would  strictly speaking be unrealistic. 
Nonetheless, this model is sufficient and convenient for our purposes to illustrate how induced GWs depend on non-gaussianity, when the PBH formation probability is fixed to account for the entire or part of dark matter. Other examples of non-gaussianity will \textcolor{black}{also} be  discussed below. 
}:
\begin{equation}
P(\zeta)=\frac{1}{2\sqrt{2}\tilde{\sigma} \Gamma\left(1+1/p\right)}\exp \left[-\left(\frac{|\zeta |}{\sqrt{2}\tilde{\sigma}}\right)^p\right],\label{pdf2}
\end{equation}
where $\tilde{\sigma}$ and $p$ are positive. 
This function satisfies $\int_{-\infty}^\infty P(\zeta)d\zeta =1$ and reduces to a gaussian PDF when $p=2$. 
When $p<2$, the PDF decreases more slowly than the gaussian case: that is, the high-sigma tail is enhanced. 
The dispersion of $\zeta$ is 
\begin{equation}
\sigma^2\equiv \int_{-\infty}^\infty \zeta^2 P(\zeta)d\zeta=\frac{2\Gamma(1+3/p)}{3\Gamma(1+1/p)}\tilde{\sigma}^2,\label{sigma}
\end{equation}
where $\Gamma(a)$ is a gamma function. 
In particular, $\sigma=\tilde{\sigma}$ when $p=2$, as it should be. 
The abundance of PBHs is
\begin{equation}
\beta=\int_{\zeta_{\mathrm{th}}}^\infty P(\zeta)d\zeta =\frac{\Gamma(1/p, 2^{-p/2}(\zeta_{\mathrm{th}}/\tilde{\sigma})^p)}
{2p\Gamma(1+1/p)},
\end{equation}
where $\zeta_{\mathrm{th}}$ is the threshold for the PBH formation\footnote{The threshold for PBH formation has been discussed in the literature \cite{Carr:1975qj,Harada:2013epa}, and strictly speaking the formation condition depends on the perturbation profile and hence cannot be described by a single parameter \cite{Shibata:1999zs,Polnarev:2006aa,Nakama:2013ica}.}, for which we take $\zeta_{\mathrm{th}}=0.5$, and $\Gamma(a,z)$ is an incomplete gamma function. 
This can be solved for $\tilde{\sigma}$ as 
\begin{equation}
\tilde{\sigma}=\frac{2^{-1/2}\zeta_{\mathrm{th}}}{Q^{-1}(1/p,2\beta)^{1/p}},\label{sigmatilde}
\end{equation}
where $Q^{-1}(a,z)$ is the inverse of the regularized incomplete gamma function $Q(a,z)\equiv \Gamma(a,z)/\Gamma(a)$, 
namely, $z=Q^{-1}(a,s)$ if $s=Q(a,z)$.
Let us consider the scenario of Ref.~\cite{Bird:2016dcv} and fix $\beta$ to account for the entire dark matter by PBHs. The initial abundance $\beta$ is related to the current abundance $\Omega_{\mathrm{PBH}}$, normalized by the current critical density, as
\begin{equation}
\beta= \frac{\rho_{\mathrm{PBH}}}{\rho_r}=\left(\frac{g}{g_0}\right)^{1/3}a\frac{\Omega_{\mathrm{PBH}}}{\Omega_r}.
\end{equation}
By eliminating the scale factor via
\begin{equation}
M=\frac{\gamma}{2G}\left(\frac{3}{8\pi G\rho_r}\right)^{1/2}
=\gamma\Omega_r^{-1/2}a^2M_0\left(\frac{g}{g_0}\right)^{1/6},
\end{equation}
we find
\begin{equation}
\beta=\gamma^{-1/2}\left(\frac{g}{g_0}\right)^{1/4}
\frac{\Omega_{\mathrm{PBH}}}{\Omega_r^{3/4}}\left(\frac{M}{M_0}\right)^{1/2}\simeq 1.2\times 10^{-8}\gamma^{-1/2}\left(\frac{g}{10.75}\right)^{1/4}\left(\frac{\Omega_{\mathrm{PBH}}}{0.3}\right)\left(\frac{M}{30M_\odot}\right)^{1/2}.\label{beta}
\end{equation} 
When $\beta$ is fixed, $\sigma$ (or the ratio $\zeta_{\mathrm{th}}/\sigma$) is smaller (larger) for a smaller value of $p$, as is shown in the top panel of Fig.~\ref{main}, 
which can be obtained from Eqs.~(\ref{sigma}) and (\ref{sigmatilde}). See also Fig.~7 of Ref.~\cite{Nakama:2016kfq}, which shows PDFs with different values of $p$ for a fixed $\beta$. 
Since induced GWs are roughly estimated by $\Omega_{\mathrm{gw}}(f)\sim \sigma^4\Omega_r$ \cite{Saito:2009jt}\footnote{
\textcolor{black}{The peak amplitude of induced GWs depends  on the spectrum of the curvature perturbations; if the width of the latter is narrow, the peak amplitude is larger than this simple estimation \cite{Saito:2009jt}}.
},
they are smaller (larger) when $p<2$ ($p>2$) than the gaussian case when $\beta$ is fixed. 
This is very similar to what was described in Ref.~\cite{Nakama:2016kfq}, which sought to evade constraints from spectral distortions of the cosmic microwave background while producing massive PBHs to account for supermassive black holes found at high redshifts. In the two-field inflation models discussed there, the RMS was kept sufficiently small so that the dissipation of global fluctuations does not produce substantial spectral distortions. 
The dependence of induced GWs on $p$ is shown in the center panel of Fig.~\ref{main}, along with other issues described in the following.

Although we have discussed induced GWs from global fluctuations, determined by the RMS amplitude,  the
locally largest GWs due to generation on superhorizon scales and shock formation are expected where high-sigma peaks collapse to form PBHs. 
Another related mechanism is GWs due to non-spherical collapse (see also Ref.~\cite{1980A&A....89....6C}). 
These GWs are locally largest but are suppressed overall due to the extreme rareness of collapsing peaks. 
The contributions of the three mechanisms would not be hugely different, and hence 
we focus on GWs from non-spherical collapse as an example. 
It will be convenient to write $\Omega_{\mathrm{gw}}\sim f \beta \Omega_r\sim 10^{-12}f$, where $f<1$ is the 
efficiency factor or the ratio of the energy of GWs to the mass of PBHs. The precise determination of $f$ would be challenging, but even for the maximal case of $f=1$, these GWs are smaller than what are expected from supermassive black holes, and hence the precise determination of the efficiency $f$ is not crucial for our discussions. With this in mind, 
the efficiency $f$ may roughly be given by the square of ellipticity $\epsilon$ of collapsing overdensities (see e.g. \cite{Grishchuk:2000gh}), and the ellipticity of high-sigma peaks may be $\epsilon\sim 0.1$,\footnote{
Higher-sigma peaks, relevant to PBH formation, tend to be more spherically symmetric at least for a gaussian random field \cite{Bardeen:1985tr}, but still they cannot be completely spherically symmetric and estimating $\epsilon\sim 0.1$ may be reasonable. 
Though part of Ref.~\cite{Bardeen:1985tr} was recently extended to a chi-squared field in Ref.~\cite{Bloomfield:2016civ}, 
to what extent this tendency about sphericity of peaks holds for  non-gaussian random fields is uncertain. } then $f\sim 0.01$. What is shown in the center panel of Fig.~\ref{main} corresponds to this value as an example.

In Ref.~\cite{Lentati:2015qwp},
new limits are provided on stochastic GWs based on
observations of six pulsars for 18 years by the European Pulsar Timing Array.
They show constraints on the GW spectrum assuming some power-law spectrum, and also constraints at a series of discrete frequencies without assuming a power-law, which is more suited for our purposes. The lowest frequency constrained there is $(18\:\mathrm{years})^{-1}\simeq 1.8\times 10^{-9}$Hz, at which $\Omega_{\mathrm{gw}}\lesssim 3\times 10^{-9}$ (their Fig.~1), which is shown in the center panel of Fig.~\ref{main}. This frequency approximately corresponds to the PBH mass of $\sim 30M_\odot$ from (\ref{freq})\footnote{
This constrained mass is somewhat larger than what is given in \cite{Bugaev:2010bb}, i.e., $(0.03-10)M_\odot$. This is thanks to more recent data based on longer observations used in \cite{Lentati:2015qwp}. 
Future data would probe induced GWs associated with the formation of even more massive PBHs: from (\ref{massk}) the PBH mass is related to the observation time $T$ as
\begin{equation}
M=98\,\gamma M_\odot\left(\frac{g}{10.75}\right)^{-1/6}\left(\frac{T}{50\,\mathrm{years}}\right)^2.
\end{equation}
}, noting that the width of the GW spectrum is expected to be on the same order as the peak frequency \cite{Saito:2009jt}.\footnote{
When finalizing our work, we became aware of \textcolor{black}{Refs.~\cite{Inomata:2016rbd,Orlofsky}, which also discuss} induced GWs associated with the formation of solar-mass PBHs \textcolor{black}{assuming gaussian fluctuations}, potentially probed by pulsar timing. \textcolor{black}{In Ref. \cite{Inomata:2016rbd} it was noted} that if $\gamma$, introduced in Eq.~(\ref{mass}) of this paper, is $\simeq 0.2$, obtained in Ref.~\cite{Carr:1975qj}, and in addition if the GW spectrum is sufficiently steep, then pulsar timing constraints can be evaded, while this  is no longer the case if $\gamma$ is larger, say $\gamma\simeq 1.$ 
\textcolor{black}{Similarly, in Ref. \cite{Orlofsky} the spectra of induced GWs for several models and their observability by pulsar timing were discussed.}
In this paper we discuss a different aspect of the problem: i.e., the effects of the non-gaussianity of primordial fluctuations. 
}

We also show a predicted sensitivity for the Square Kilometer Array of $\Omega_{\mathrm{gw}}\sim 10^{-12}$ as well as that of stochastic GWs from binary supermassive black holes of $\Omega_{\mathrm{gw}}\sim 10^{-10}$ shown in Ref.~\cite{Kramer:2010tm}, where the latter is based on Refs.~\cite{Sesana:2008mz,Sesana:2010mx}.

Though our estimates of induced GWs are crude, those for the case of gaussian fluctuations may already be inconsistent with the existing upper limit from pulsar timing. In addition, some types of models with non-gaussianity that reduce the ratio $\zeta_{\mathrm{th}}/\sigma$ are also inconsistent. 
 A moderate amount of non-gaussianity seems to suffice  to evade the existing upper limit, since the upper limits and induced GWs for the gaussian case are comparable. Depending on the amount of non-gaussianity, induced GWs can still be larger than those from supermassive black hole binaries. Even if in the future we were to observe stochastic GWs that appear to be those from supermassive black hole binaries, we cannot fully exclude the possibility of a PBH dark matter scenario,  since in principle the amplitude of the induced GWs associated with PBH formation can be made smaller depending on non-gaussianity, as shown in the center panel of Fig.~\ref{main}. Current/future data from pulsar timing should provide restrictions on models which predict PBHs of tens of solar masses.

The dependence on $\beta$ of induced GWs is relatively weak (logarithmic), since the dependence of $\beta$ on the RMS, which determines induced GWs, is exponential. 
See the bottom panel of Fig.~\ref{main}, which corresponds to its center panel with different values of $\beta$. 
This shows that induced GWs are also useful to constrain scenarios where PBHs constitute only part of the entire dark matter, such as the scenario of Ref.~\cite{Sasaki:2016jop}.
In contrast local GWs are proportional to $\beta$.

\begin{figure}[tp]
\begin{center}
\includegraphics[width=13cm,keepaspectratio,clip]{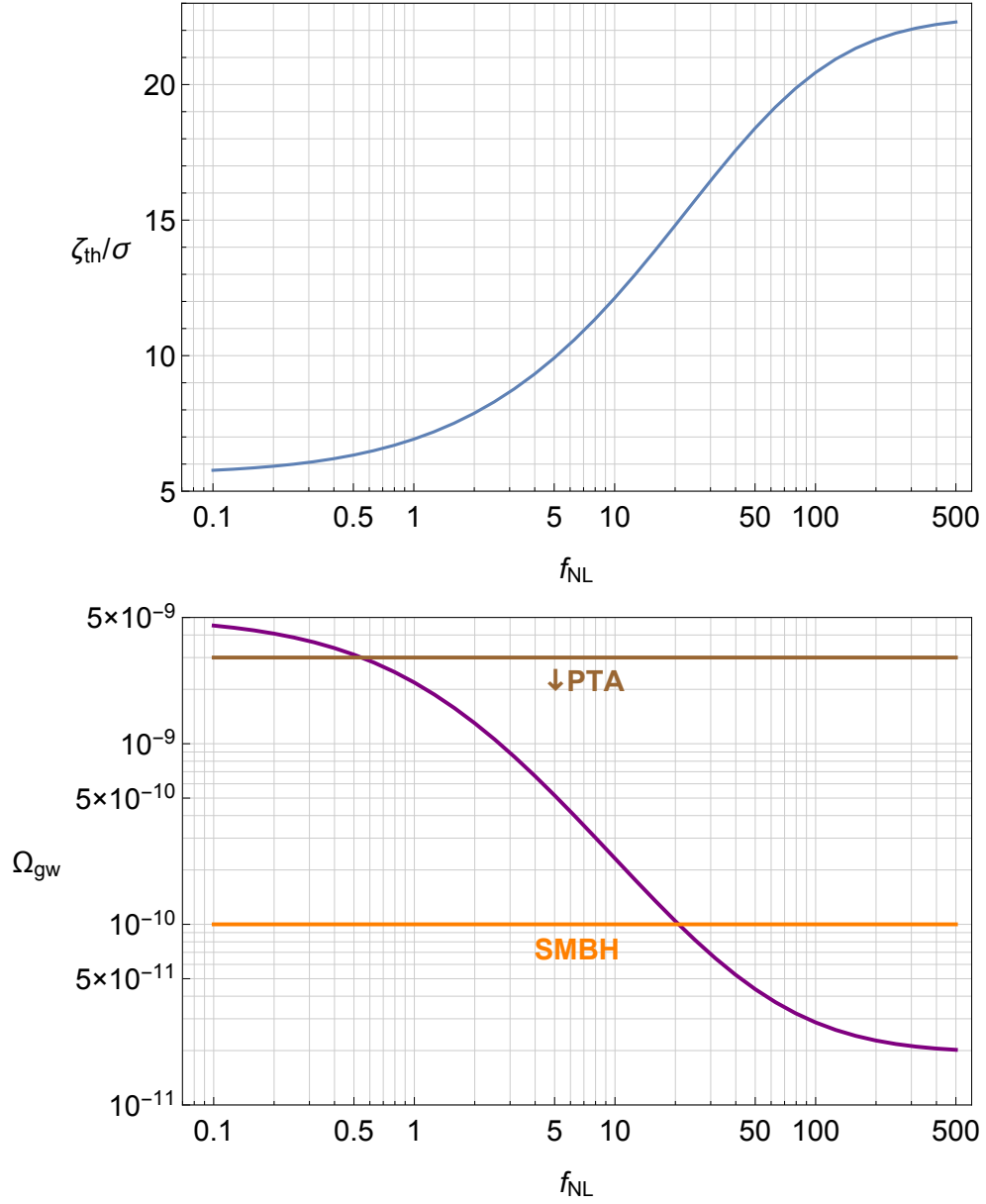}
\end{center}
\caption{
Top: The dependence of the ratio $\zeta_{\mathrm{th}}/\sigma$ on $f_{\mathrm{NL}}$ of Eq.~(\ref{fnl}). Bottom: The dependence of induced GWs on $f_{\mathrm{NL}}$.
}
\label{fnlfig}
\end{figure}

\begin{figure}[tp]
\begin{center}
\includegraphics[width=13cm,keepaspectratio,clip]{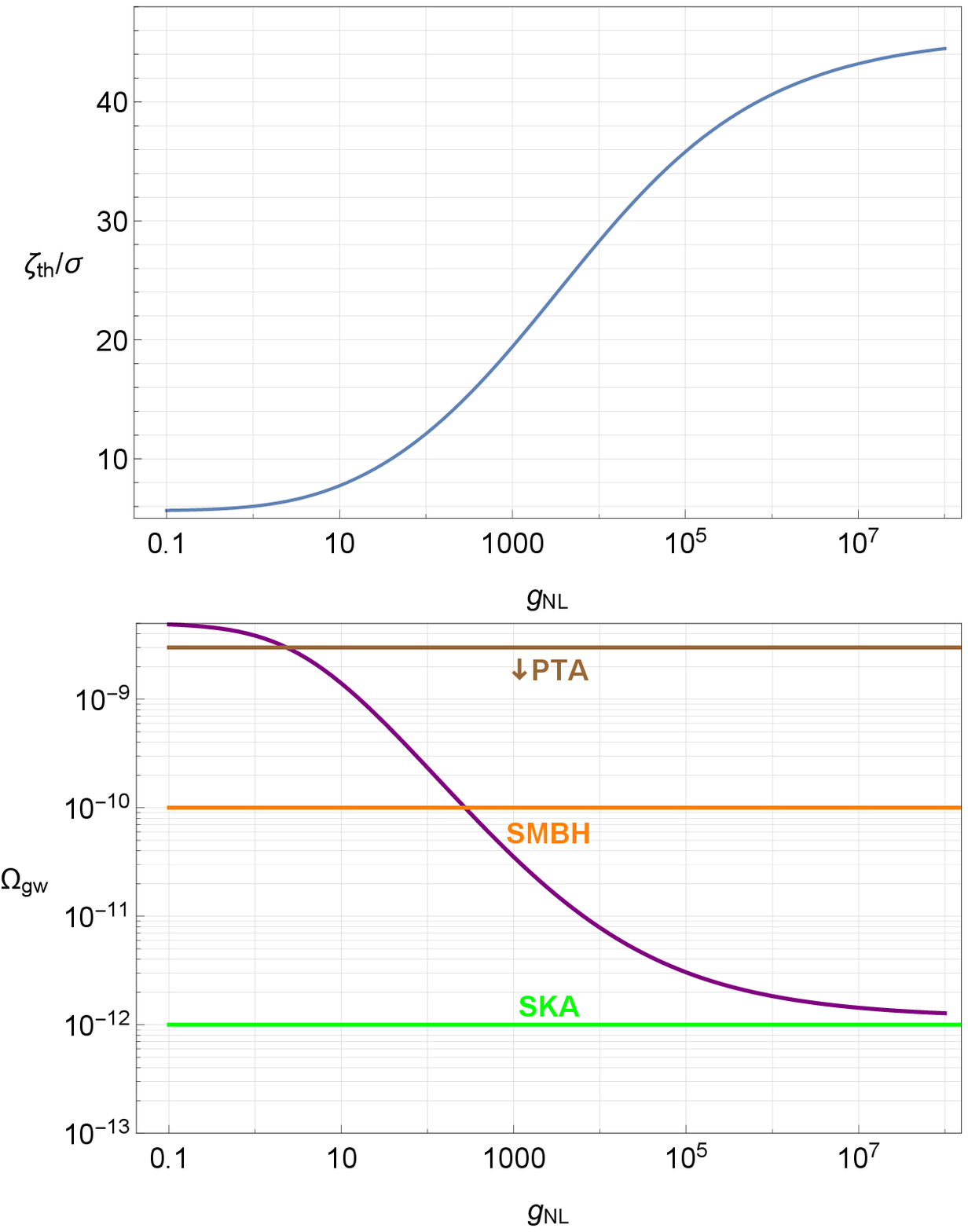}
\end{center}
\caption{
Top: 
The dependence of the ratio $\zeta_{\mathrm{th}}/\sigma$ on $g_{\mathrm{NL}}$ of Eq.~(\ref{gnl}). 
Bottom: The dependence of induced GWs on $g_{\mathrm{NL}}$.
}
\label{gnlfig}
\end{figure}

One may also consider other types of non-gaussianity.
For instance let us consider the following local models of non-gaussianity, used in Ref.~\cite{Byrnes:2012yx}, restricting $f_{\mathrm{NL}},g_{\mathrm{NL}}>0$:
\begin{equation}
\zeta=\zeta_G+\frac{3}{5}f_{\mathrm{NL}}(\zeta_G^2-\sigma_G^2),\label{fnl}
\end{equation}
\begin{equation}
\zeta=\zeta_G+g\zeta_G^3,\quad g\equiv \frac{9}{25}g_{\mathrm{NL}},\label{gnl}
\end{equation}
where $\zeta_G$ is gaussian and its variance is $\langle\zeta_G^2\rangle=\sigma_G^2.$
Since $\langle\zeta_G^4\rangle=3\sigma_G^4$
and
$\langle\zeta_G^6\rangle=15\sigma_G^6$, the variance $\langle\zeta^2\rangle=\sigma^2$ of $\zeta$ for these models is
\begin{equation}
\langle\zeta^2\rangle=\sigma_G^2+2\left(\frac{3}{5}f_{\mathrm{NL}}\right)^2\sigma_G^4,\label{varfnl}
\end{equation}
\begin{equation}
\langle\zeta^2\rangle=\sigma_G^2+6g\sigma_G^4+15g^2\sigma_G^6.\label{vargnl}
\end{equation}
In Ref.~\cite{Byrnes:2012yx} the dependence of the PBH upper limits on the power spectrum of the primordial curvature perturbations on non-gaussianity was investigated. They found that upper limits are tighter for larger positive $f_{\mathrm{NL}}$ and $g_{\mathrm{NL}}$. This is because the high-sigma tail is larger for these cases, and hence the variance of $\zeta$ is smaller for a fixed $\beta$. Likewise, induced GWs are smaller when fixing $\beta=10^{-8}$ for $f_{\mathrm{NL}},g_{\mathrm{NL}}>0$.
Like Ref.~\cite{Byrnes:2012yx} we are also interested in regimes where the non-gaussian terms of Eq.~(\ref{fnl}) and (\ref{gnl}) are no longer small corrections to the gaussian part and can dominate. 
We follow Ref.~\cite{Byrnes:2012yx} to calculate the abundance of PBHs for the above models of non-gaussianity\footnote{
The signs of the square roots of Eq.~(37) of Ref.~\cite{Byrnes:2012yx} should probably be flipped. }. 
When we fix $\beta=10^{-8}$, $\sigma$ is determined, which then determines induced GWs, but this time it depends on $f_{\mathrm{NL}}$ or $g_{\mathrm{NL}}$. 
The dependence of the ratio of the threshold to the RMS and induced GWs on $f_{\mathrm{NL}},g_{\mathrm{NL}}$ for $\beta=10^{-8}$ are shown in Figs.~\ref{fnlfig} and \ref{gnlfig}. 

The behavior shown in the top panel of Fig.~\ref{fnlfig} can be understood as follows. 
From the plot one can confirm that when $f_{\mathrm{NL}}\simeq 65$ ($\zeta_{\mathrm{th}}/\sigma\simeq 19.2$) the two terms in Eq.~(\ref{varfnl}) are comparable, 
and this value roughly coincides with the inflection point. 
When $f_{\mathrm{NL}}$ is sufficiently larger than this value, the first gaussian term becomes negligible. 
In this case the parameter $f_{\mathrm{NL}}$ becomes no longer meaningful since redefining $(3f_{\mathrm{NL}}/5)^{1/2}\zeta_G\rightarrow \zeta_G$, 
$\zeta$ is written as $\zeta_G^2-\sigma_G^2$, and the PDF of $\zeta$ is a $\chi^2$ distribution and is controlled by a single parameter $\sigma_G$. 
This explains the asymptotic behavior of the top panel of Fig.~\ref{fnlfig}. In the large $f_{\mathrm{NL}}$ limit the abundance is calculated as
 $\beta=\mathrm{erfc}[(\tilde{\zeta}_{\mathrm{th}}/2)^{1/2}]$, 
where  $\mathrm{erfc}(x)$ is the complementary error function and $\tilde{\zeta}_{\mathrm{th}}=\sqrt{2}\zeta_{\mathrm{th}}/\sigma+1$. Setting $\beta=10^{-8}$ this relation can be solved for the ratio as $\zeta_{\mathrm{th}}/\sigma\simeq 22.5$, which approximately corresponds to the asymptotic value read off from the top panel of Fig.~\ref{fnlfig}. 
The bottom panel of Fig.~\ref{fnlfig} shows the corresponding dependence of induced GWs on $f_{\mathrm{NL}}$. Pulsar-timing constraints can be evaded depending on $f_{\mathrm{NL}}$, and when $f_{\mathrm{NL}}$ is sufficiently large, induced GWs can also be smaller than or comparable to any prediction for supermassive-black-hole binaries.

The case of cubic non-gaussianity shown in Fig.~\ref{gnlfig} can also be understood similarly. 
When $g_{\mathrm{NL}}\simeq 2000$, the first and second terms of Eq.~(\ref{vargnl}) become comparable, and then the second and third terms become comparable when
 $5g^{1/2}\zeta_{\mathrm{th}}/[4\sqrt{3}(\zeta_{\mathrm{th}}/\sigma)]\simeq 1$ or $g_{\mathrm{NL}}\simeq 2\times 10^4$, which roughly corresponds to the inflection point. In the large $g_{\mathrm{NL}}$ limit, redefining $g^{1/3}\zeta_G\rightarrow \zeta_G$, $\zeta=\zeta_G^3$, and the PDF of $\zeta$ as well as the abundance $\beta$ in this case is
 \begin{equation}
P(\zeta)=\frac{d\zeta^{1/3}}{d\zeta}\frac{1}{\sqrt{2\pi}\sigma_G}\exp\left[-\frac{\zeta^{2/3}}{2\sigma_G^2}\right]=
\frac{\zeta^{-2/3}}{3\sqrt{2\pi}\sigma_G}\exp\left(-\frac{\zeta^{2/3}}{2\sigma_G^2}\right),
\end{equation}
\begin{equation}
\beta=\frac{1}{2}\mathrm{erfc}\left[\frac{1}{\sqrt{2}}\left(\frac{\zeta_{\mathrm{th}}}{\sigma/\sqrt{15}}\right)^{1/3}\right].
\end{equation}
Once more, setting $\beta=10^{-8}$ we can solve for the ratio as $\zeta_{\mathrm{th}}/\sigma\simeq 45.5$, which approximately gives the asymptotic value shown in the top panel of Fig.~\ref{gnlfig}.
For the case of cubic non-gaussianity the ratio $\zeta_{\mathrm{th}}/\sigma$ can be made larger, and hence $\Omega_{\mathrm{gw}}$ can be made smaller than quadratic non-gaussianity.

One may also consider other types of non-gaussianity, such as those investigated in Ref.~\cite{Young:2013oia}. Higher order terms are expected to make $\zeta_{\mathrm{th}}/\sigma$ even larger, and hence $\Omega_{\mathrm{gw}}$ smaller. The PDF of Eq.~(\ref{pdf2}) would probably correspond to the PDF of $\zeta$ when it is written as some power of a gaussian variable, i.e., $\zeta\sim \zeta_G^{2/p}$, very roughly. 
\textcolor{black}{For comparison, the PDF of $\zeta=\zeta_G^{2i-1},(i=1,2,3,\cdots)$ is
\begin{equation}
P(\zeta)=\frac{\zeta^{-2(i-1)/(2i-1)}}{(2i-1)\sqrt{2\pi}\sigma_G}\exp\left[-\frac{\zeta^{2/(2i-1)}}{2\sigma_G^2}\right],
\end{equation}
and the PDF of $\zeta=\zeta_G^{2i}-\langle\zeta_G^{2i}\rangle$ is
\begin{equation}
P(\zeta)=\frac{(\zeta+\langle\zeta_G^{2i}\rangle)^{-(2i-1)/2i}}{i\sqrt{2\pi}\sigma_G}\exp\left[-\frac{(\zeta+\langle\zeta_G^{2i}\rangle)^{1/i}}{2\sigma_G^2}\right].
\end{equation}
}

In addition, two-field-inflation models were recently discussed  in Ref.~\cite{Nakama:2016kfq}, 
which can predict PBHs with a peaked mass spectrum while keeping the RMS sufficiently small. In this model, only a tiny fraction of patches of some radius inside the current observable universe experience more expansion than elsewhere during inflation, and consequently these patches can have large curvature perturbations and later collapse to form PBHs during the radiation domination. Except for the presence of these rare patches with large curvature perturbations, primordial fluctuations are almost scale invariant, hence the small RMS of $\sigma^2\sim 10^{-9}$. There a possibility of  generating PBHs massive enough to explain supermassive black holes without causing large spectral distortions of the cosmic microwave background was discussed, but in principle producing PBHs of tens of solar masses without generating large induced GWs should also be possible. 
Since $\sigma^2\sim 10^{-9}$,
induced GWs in that model are roughly estimated to be  $\Omega_{\mathrm{gw}}\sim 10^{-22}$.


\section{Discussion}
The formation of PBHs of $\sim 30\, M_\odot$ is likely to be associated with stochastic GWs of interesting frequencies and amplitudes. 
This is partly because the RMS amplitudes of primordial fluctuations may be relatively large, and in this case substantial GWs are generated at second order in scalar perturbations mostly before  horizon reentry.  For instance, if we assume the full PBH dark matter scenario of \cite{Bird:2016dcv} and gaussianity of primordial fluctuations, these induced GWs are comparable to the current limits from pulsar timing arrays. Naturally, the power-spectrum amplitude and hence induced-GW amplitude depend on non-gaussianity, and we have investigated that dependence adopting examples of non-gaussianity from Refs.~\cite{Byrnes:2012yx,Nakama:2016kfq}. Depending on non-gaussianity, current limits from pulsar timing can be evaded, and also induced GWs can be larger or smaller than  predictions for stochastic GWs from supermassive-black-hole binaries. \textcolor{black}{Predicted frequency spectra for induced GWs \cite{Saito:2008jc,Saito:2009jt} and stochastic GWs from supermassive black holes \cite{Lentati:2015qwp,Kramer:2010tm,Sesana:2008mz,Sesana:2010mx} are in general different and hence they could be distinguished in future experiments such as the Square Kilometer Array, in principle.}
In the extreme example of Ref.~\cite{Nakama:2016kfq}, the power-spectrum amplitude is even as small as the standard almost-scale-invariant spectrum seen on large scales, while generating an interesting amount of PBHs. 
Existing/future data from pulsar timing will provide important restrictions on models which predict the formation of PBHs of tens of solar masses. 
It would be worthwhile to regard constraints on the abundance of PBHs from induced GWs as a conditional constraint, or more  generally,  induced GWs provide joint constraints on the abundance of PBHs and non-gaussianity, similarly to the fact that the absence of PBHs provides joint constraints on the power spectrum of  primordial curvature perturbation and non-gaussianity \cite{Byrnes:2012yx}.

\textcolor{black}{
Though we have considered non-gaussian primordial fluctuations on small scales, the mechanism causing small-scale fluctuations relevant to PBH formation and that causing large-scale fluctuations have to be sufficiently decoupled, or PBH-scale and large-scale fluctuations are sufficiently uncorrelated, so as not to generate unacceptably large PBH dark matter isocurvature perturbations on large scales \cite{Tada:2015noa,Young:2015kda}.
The running-mass and axion-curvaton model were noted to be unaffected by this constraint \cite{Carr:2016drx}. 
Note that $f_{\mathrm{NL}}$ and $g_{\mathrm{NL}}$ of this paper can be irrelevant to those constrained by this isocurvature effect \cite{Byrnes:2009pe,Tada:2015noa,Carr:2016drx}, or those constrained by the Planck collaboration \cite{Ade:2015ava}. 
One may also consider a burst of particle production associated with some field coupled to the inflaton during inflation, well after large-scale fluctuations are generated by quantum fluctuations of the inflaton, causing non-gaussian, large-amplitude fluctuations on particular scales due to nonlinear interactions \cite{Barnaby:2009mc,Barnaby:2009dd,Barnaby:2010ke,Barnaby:2010sq,Erfani:2015rqv,Cheng:2016qzb}. 
In addition, Ref. \cite{Bullock:1996at} considered  plateau-type features in the inflaton potential, which also lead to a spike in the primordial spectrum, 
but the properties of non-gaussianity in those models have been debated \cite{Ivanov:1997ia,Bullock:1998mi}. 
}

When PBHs constitute only part of the entire dark matter, they are likely to be associated with a copious production of dark matter minihalos as well \cite{Kohri:2014lza}, which may potentially be excluded by the absence of associated annihilation signals \cite{Bringmann:2011ut}  or their gravitational or dynamical effects. Nevertheless, an overproduction of minihalos can also be evaded by non-gaussianity.

Higher-frequency global GWs associated with PBHs smaller than a solar mass can also be probed by space-based and ground-based GW detectors  \cite{Saito:2008jc,Saito:2009jt,Bugaev:2009zh,Bugaev:2010bb}. 
See also \cite{Kuroda:2015owv} for sensitivity curves of various experiments. 
Our discussions naturally apply to different masses of PBHs, that is, constraints on them from induced GWs depend on primordial non-gaussianity. 
In addition,
lower-frequency GWs 
may also be probed  in the future by lensing of 21-cm fluctuations  \cite{Book:2011dz}

\section*{ACKNOWLEDGMENTS}
We thank Teruaki Suyama for helpful discussions.  T. N.  was partially supported by a  JSPS Postdoctoral Fellowship for Research Abroad. J. S. was supported by the ERC Project No. 267117 (DARK) hosted by Universite Pierre et Marie Curie (UPMC) - Paris 6.  This work was supported at Johns Hopkins by the Simons Foundation, NSF Grant No. PHY-1214000, and NASA ATP Grant No. NNX15AB18G. 
\bibliographystyle{h-physrev}
\bibliography{ref}
\end{document}